\documentclass[aps,nofootinbib,superscriptaddress, showpacs,preprintnumbers,  nofootinbibt,twocolumn]{revtex4-2}
\usepackage{eurosym}
\usepackage{amsmath}
\usepackage{bm}
\usepackage{amsfonts}
\usepackage{amssymb}
\usepackage{graphicx}
\usepackage{mathtools}
\usepackage{color}
\usepackage{physics}

\def\be{\begin{equation}}
\def\ee{\end{equation}}
\def\bea{\begin{eqnarray}}
\def\eea{\end{eqnarray}}

\begin{document}

\title{The Stochastic-Dissipative St\"{o}rmer Problem-Trajectories and Radiation Patterns}
\author{Tiberiu Harko}
\email{tiberiu.harko@aira.astro.ro}
\affiliation{Department of Physics, Babeș-Bolyai University, Kogălniceanu Street 1, 400084 Cluj-Napoca,   Romania}
\affiliation{Astronomical Institute, Romanian Academy of Sciences, Cluj-Napoca Branch, 19 Ciresilor Street,  400487 Cluj-Napoca, Romania}
\author{Gabriela Raluca Mocanu}
\email{gabriela.mocanu@aira.astro.ro}
\affiliation{Astronomical Institute, Romanian Academy of Sciences, Cluj-Napoca Branch, 19 Ciresilor Street,  400487 Cluj-Napoca, Romania}

\begin{abstract}
We consider a generalization of the classical nonrelativistic St\"{o}rmer problem, describing the motion of charged particles in a purely magnetic dipole field, by taking into account the effects of the dissipation, assumed to be of friction type, proportional to the velocity of the particle, and of the presence of stochastic forces. In the presence of dissipative/stochastic effects, the  motion of the particle in the magnetic dipole field can be described by a generalized Langevin type equation, which generalizes the standard Lorentz force equation. We perform a detailed numerical analysis of the dynamical behavior of the particles in a magnetic dipolar field in the presence of dissipative and stochastic forces, as well as of the electromagnetic radiation patterns emitted during the motion. The effects of the dissipation coefficient and of the stochastic force on the particle motion and on the emitted electromagnetic power are investigated, and thus a full description of the spectrum of the magnetic dipole type electromagnetic radiation and of the physical properties of the motion is also obtained. The power spectral density of the emitted electromagnetic power is also obtained for each case, and, for all considered St\"{o}rmer type models, it shows the presence of peaks in the radiation spectrum, corresponding to certain intervals of the frequency.

\end{abstract}

\maketitle
\color{blue}
\tableofcontents

\color{black}

\section{Introduction}

Magnetic fields are a common occurrence  on all astrophysical, and even cosmological scales, their presence being detected from the galactic or extragalactic level \citep{M1,M2} to stars, Sun, planets, and Earth  \citep{M3}. At its surface the magnitude of the Earth's magnetic field varies from around 0.25 to 0.65 Gauss.  It can be approximated as the field of a magnetic dipole tilted at an angle of around 11 degrees with respect to the rotational axis of the Earth  \cite{EMF}. For the Earth dipolar field the dipole momentum is
$M_{z}=7.9\times 10^{25}\;\mathrm{G\; cm^{3}}=7.9\times 10^{15}\;\mathrm{T\;
m^{3}}$.

 Hence the study of the motion of charged particles in magnetic fields is of major theoretical, as well as observational and practical importance. One of the early landmark investigations, with important applications, were the studies by St\"{o}rmer \cite{St1,St2,St3,St4,St4a, St5,St6,St7} of dynamics of particles in a purely magnetic dipolar field.  St\"{o}rmer's analysis was mainly motivated by finding an explanation of the Northern Lights, but the model as well as the obtained results did find many applications in astronomy and astrophysics, among the most interesting ones being the explanation
of the dynamics of electrons or ions that are present in the radiation belts formed in the planetary magnetic fields, as initiated in \cite{B1,B2}.

The {\it St\"{o}rmer problem} can be formulated as follows. Consider a non-relativistic particle of mass $m$, charge $q$, and momentum $\vec{p}$ in motion in the magnetic dipolar field of a magnetic moment $\vec{M}$. The Hamiltonian of the system is
\be
H=\frac{1}{2m}\left(\vec{p}-\frac{q}{c}\vec{A}\right)^2,
\ee
where  the vector potential $\vec{A}$ is given by $\vec{A}=\left(\vec{M}\times \vec{r}\right)/r^3$, and $r=\sqrt{x^2+y^2+z^2}$, respectively  \citep{B2}. By choosing the direction of $\vec{M}$ along the $z$ axis we have $\vec{M}=(0,0,M)$, and, by taking $m=1$, and denoting $a=qM/c$, the Hamiltonian can be obtained in the form
\be
H=\frac{1}{2}\left[\left(p_x+\frac{ay}{r^3}\right)^2+\left(p_y-\frac{ax}{r^3}\right)^2+p_z^2\right],
\ee
or, equivalently,
\be\label{H1}
H=\frac{1}{2}\left(p_x^2+p_y^2+p_z^2\right)+\frac{a}{r^3}\left(yp_x-xp_y\right)+\frac{a^2}{2r^6}\left(x^2+y^2\right).
\ee

Together with the Hamiltonian (\ref{H1}), the projection of the angular momentum in the direction of $\vec{M}$, $L_z=xp_y-yp_x$ is also an integral of the motion. 

We call the problem of finding the solutions, and physical properties, of the equations of motion derived from the Hamiltonian (\ref{H1}) the {\it Classical St\"{o}rmer problem} (CSP). In \cite{Int} it was shown, by using the Ziglin-Yoshida method that the CSP {\it is non-integrable}.  An important result of the CSP is the proof of the existence of allowed and forbidden regions for charged particles, as well as of  particle storage regions.

The existence of trapping regions is the theoretical basis for the understanding of the particle capture and the formation of the radiation belts of the Earth and of other planets \cite{B1,B2}. Even in its simplified formulation, the St\"{o}rmer model offers a good description, at least on a qualitative level, of the three main physical effects observed in the Earth magnetosphere: the Van Allen radiation belts \cite{B1,B2}, the polar aurora \cite{St4,St5,St6}, and the South Atlantic anomaly, a region where the Earth's inner radiation belt is closest to the surface of the Earth, leading to an increased flux of energetic particles \cite{An}.  The South Atlantic anomaly is located at a height of around 200 km, or at $0.031R_{\oplus}$, and it exposes orbiting satellites (including the International Space Station) to high levels of ionizing radiation.

The St\"{o}rmer problem has been generalized for many other physical configurations.
The St\"{o}rmer problem for the motion of a charged particle in the field of rotating uniformly magnetized celestial body is called the {\it Rotational St\"{o}rmer Problem} (RSP). In \cite{Schust} the charged particle trapping in the electromagnetic field of the parallel
rotator was investigated. 

By considering an electromagnetic field with azimuthal symmetry represented by the
electric potential $A_0=A_0(R,z)$ superimposed on the dipolar magnetic field, the Lagrangian of the particle can be written as
\be
L=-mc^2\sqrt{1-\frac{v^2}{c^2}}+\frac{e\mu}{c}\frac{R^2\dot{\phi}}{\left(R^2+z^2\right)^{32}}-eA_0.
\ee
It follows that within the considered electromagnetic field
configuration, two or even three disconnected torus-shaped trapping
regions may exist.

A systematic study of the rotational St\"{o}rmer problem (RSP), with the electric
field due to the rotation of the body included, of the {\it gravitational St\"{o}rmer problem (GSP)}, with only the Keplerian gravity considered
and with the effects of the corotational electric field neglected, and of the full system (RGSP), including both electric and gravitational fields, was performed in \cite{How} and \cite{Dull}, respectively. In the presence of a potential $U(r)$ generating electric and gravitational forces the equation of motion of a charged particle is given by
\be
m\frac{d^2{\vec{r}}}{dt^2}=\frac{q}{c}\vec{v}\times \vec{B}-\nabla U\left(\vec{r}\right).
\ee

By assuming that the magnetosphere is a highly conducting plasma, and it may  be
assumed to corotate rigidly with the planet with uniform angular velocity $\Omega$, the electric field in the inertial frame is given by
\be
\vec{E}=-\frac{1}{c}\left(\boldsymbol{\Omega}\times \vec{r}\right)\times \vec{B}.
\ee

The electric field can be described in terms of a stream function $\Psi=\left(x^2+y^2\right)/r^3$, so that $q\vec{E}=-\gamma \Omega \nabla \psi$, where $\gamma =qM/c$ \citep{Dull}. Hence the potential describing the particle in the combined gravitational and electric fields is  obtained as
\be
U\left(\vec{r}\right)=-\sigma _g\frac{GM_p m}{r}+\sigma _{\gamma}\gamma \Omega \Psi,
\ee
where $M_p$ is the mass of the planet, and $\sigma _g$ and $\sigma _{\gamma}$ are two parameters describing the strength of the gravitational and electric forces, respectively. The inertial frame Hamiltonian of the problem is given in cylindrical coordinates $(\rho,\phi, z)$ by \citep{How}
\be
H=\frac{1}{2m}\left(p_{\rho}^2+p_z^2\right)+\frac{1}{2m\rho ^2}\left(p_{\phi}-\frac{q}{c}\Psi\right)^2+U+\frac{q\Omega}{c}\Psi,
\ee
where $U$ is the gravitational potential. The inclusion of the gravitational force leads to stable circular orbits, located  in a
plane situated above/below the equatorial plane of the celestial body \citep{Dull}.

The dynamical evolution of a charged particle orbiting around a rotating magnetic object was studied in \cite{In}.  The perturbation consisted of a magnetic dipole field, and a corotational electric field. The Hamiltonian of the problem is given by
\bea
H&=&\frac{1}{2m}\left(p_x^2+p_y^2+p_z^2\right)-\frac{GM_pm}{r}-\frac{\mu q}{mc}\frac{L_z}{r^3}\nonumber\\
&&+\frac{\mu ^2 q^2}{2mc^2}\frac{x^2+y^2}{r^6}+\frac{q\mu \omega }{c}\Psi.
\eea

The flow of the resulting system in the most reduced phase space was studied, and the description of
all equilibrium points and of their stability was considered. The different classes of bifurcations were also analyzed. The effect of the oblateness of the planet in the Hamiltonian function was investigated in \cite{In1}, with the
non-sphericity of the planet given by means of the $J_2$ term. The corresponding Hamiltonian function is
\bea
H&=&\frac{1}{2}\left(p_{\rho}^2+p_z^2+\frac{p_{\phi}^2}{\rho ^2}\right)-\frac{1}{r}-\delta \frac{p_{\phi}}{r^3}+\frac{\delta ^2}{2}\frac{\rho ^2}{r^6}+\delta \beta \frac{\rho ^2}{r^3}\nonumber\\
&&+3J_2\frac{z^2}{2r^5}-\frac{J_2}{2r^3},
\eea
where $r=\sqrt{\rho ^2+z^2}$, $\delta=\omega_c/w_w$, where $\omega _c$ is the cyclotron frequency, $w_k=\sqrt{M/R^3}$ is the Keplerian frequency, and $\beta =\Omega/w_k$, respectively, while $J_2$ is a dimensionless parameter, positive for an oblate planet, and negative for a prolate one.

The dynamics of a charged relativistic particle in electromagnetic field of a rotating magnetized celestial body with the magnetic axis inclined to the axis of rotation was studied in \cite{Epp0}, and the covariant Lagrangian function in the rotating reference frame was found.
The effective potential energy of the particles in the field of a rotating uniformly magnetized astrophysical object  was discussed in \cite{Epp}, with the electromagnetic field of the body represented by the superposition of a dipole magnetic
and quadrupole electric fields.  The
main difference from the classical St\"{o}rmer problem is that the single toroidal trapping
region is divided into equatorial and off-equatorial trapping regions. For other investigations of the Classical St\"{o}rmer Problem see \citep{Pina, Kol, Leg, Ersh}. 

Recently, in \cite{Ersh1},  new semi-analytical solution have been obtained for the St\"{o}rmer problem for the motion of charged particles close to the equatorial plane of Earth in the dipole magnetic field generated by a magnetized infinite cylinder. The St\"{o}rmer problem is reduced to three nonlinear ordinary differential equations of the first order, and their analytical solution was obtained in polar coordinates. 

It is the goal of the present work to introduce, and investigate, another class of St\"{o}rmer type problems, which considers the motion of a charged particle in a dipolar magnetic field in the presence of  {\it dissipative and stochastic forces}. We call this problem the {\it Stochastic-Dissipative St\"{o}rmer Problem} (SDSP). 

Its general mathematical solution is provided by the replacement of the Lorentz equation of motion, describing a deterministic particle dynamics, with a Langevin type stochastic differential equation, of the form \cite{Bal2,Bal2a,Bal3,Coff}
\begin{equation}\label{lang}
\frac{d^2\vec{r}}{dt^2}=\vec{F}\left[\vec{r}(t),\frac{d\vec{r}(t)}{dt},t\right]+\boldsymbol{\eta}(t),
\end{equation}
where $\vec{F}\left[\vec{r}(t),\frac{d\vec{r}(t)}{dt},t\right]$ gives the potential or dissipative  forces acting on the particle, while $\boldsymbol{\eta}(t)$ is a random force, modeling the stochastic physical effects generated by the cosmic environment, and by the interparticle collisions. 

Stochastic equations of Langevin type have been used to model the random oscillations of thin accretion disks in the presence of white noise \citep{HaMo1}, or of colored noise \citep{HaMo2}, respectively. The effects of the random perturbations acting on an accretion on the registered light curves were investigated in \cite{Mo1}.  The electromagnetic radiation properties of a charged non-relativistic particle in the presence of electric and magnetic fields, of an exterior non-electromagnetic potential, and of a friction and stochastic force were investigated in \cite{HaMo3}. The motion of the particle was described by a Langevin and generalized Langevin type stochastic differential equation, respectively. 

The cases of the Brownian motion with or without memory in a constant electric field, in the presence of an external harmonic potential, and of a constant magnetic field were investigated in detail. For further studies of the properties of the trajectories and radiation patterns of charged particles in constant or time-dependent magnetic fields under the influence of stochastic forces see \cite{Mo2,Mo3,Mo4}.

 From a physical point of view, the presence of a dissipative force in the St\"{o}rmer problem can be related with the interaction of the particle with the cosmic environment of the magnetosphere, which leads to interparticle collisions, which we model by means of a friction force, proportional to the particle velocity. Due to the friction forces, the energy of the particles is transferred to the medium. There are at least three physical environments (nuclear fusion devices, semiconductor devices, and planetary magnetospheres) in which the consideration of the simultaneous response of a charged particle to both collisions and variations of the  magnetic field is important. Our simplifying approach is to model collisions via a Brownian motion, or, more exactly,
as an Ornstein–Uhlenbeck process in velocity space \cite{Bal2a}. 

As a result of the presence of an external random environment, the particles gain energy from the interactions due to the random forces,
as well as from the external non-electromagnetic, electric and magnetic fields. Some of this energy is emitted in the form of the electromagnetic radiation. The electromagnetic power emitted by the particles is proportional to the square of its
acceleration, which can be computed directly from the Langevin equation describing the stochastic motion of the particles. Radiation processes are very important topics in astrophysical research, and they can provide physical mechanism to explain the emission of cosmic objects. 

The radiation mechanisms are generally the result of particle acceleration, and  plasma and collision effects must be taken into account when studying both acceleration and radiation processes.  Alternatively, the so-called jitter radiation, emitted by relativistic electrons moving in a highly nonuniform magnetic field was also investigated \cite{J1,J2,J3}. The jitter radiation of an ensemble of relativistic electrons, moving in a highly turbulent magnetic field, has a very different spectrum as compared to the standard synchrotron one. 

In the present work we investigate comparatively three distinct St\"{o}rmer type problems. The first is {\it the Classical St\"{o}rmer problem (CSP)}, in which the motion of the charged particle takes place in the presence of a pure dipole field. 

We generalize this problem by considering the effect of the dissipation (friction) on the particle motion, which leads to the {\it Classical Dissipative St\"{o}rmer Problem (CDSP)}. The dissipative force is assumed to be proportional with the velocity. 

By including an effective stochastic force into the dissipative equation of motion we arrive at the {\it Stochastic-Dissipative St\"{o}rmer Problem (SDSP)}. By neglecting the effects of the frictional force one can consider the {\it Stochastic St\"{o}rmer Problem (SSP)}, in which one considers the effects of the random force on the motion of a particle in a magnetic dipolar field. In all these cases we consider the numerical evolution of the trajectories, of the emitted electromagnetic power, and of the Power Spectral Density (PSD) of the radiation. 

The present paper is organized as follows. In Section~\ref{sect1} we introduce the basic evolution equations and physical parameters of the Stochastic-Dissipative St\"{o}rmer Problem. The numerical method for obtaining the solutions of the equations of motion is also briefly outlined. Different Classical Dissipative St\"{o}rmer and Stochastic-Dissipative St\"{o}rmer models are investigated in Section~\ref{sect2}, in which the effects of the dissipative and random forces are considered. We discuss and conclude our results in Section~\ref{sect3}.                 

\section{The Stochastic-Dissipative St\"{o}rmer Problem}\label{sect1}

The equation of motion of a nonrelativistic charged particle of mass $m$ and
charge $q$ in a magnetic field $\vec{B}$ has the standard Lorentz form,
\bea
m\frac{d^2\vec{r}}{dt^2} &=& q\frac{d\vec{r}}{dt}\times \vec{B}+\vec{F}\left[\vec{r}(t),\frac{d\vec{r}(t)}{dt},t\right] \nonumber\\
&=&q\left(\vec{v}%
\times \vec{B}\right)+\vec{F}\left[\vec{r}(t),\frac{d\vec{r}(t)}{dt},t\right] ,  \label{eq1}
\eea
where $\vec{v}=d\vec{r}/dt$ is the particle velocity, and $\vec{F}\left[\vec{r}(t),\frac{d\vec{r}(t)}{dt},t\right]$ is the external force, usually of non-magnetic origin.  In the following we use the SI system of units.

In the presence of {\it a dissipative and of a random force}, the Lorentz equation of motion can be
generalized to {\it a Langevin type stochastic differential equation},
\begin{equation}
m\frac{d^2\vec{r}}{dt^2}=q \vec{v}\times \vec{B} -\gamma m\vec{v}+m\vec{f}%
^{(s)},  \label{eq2}
\end{equation}%
where $\gamma m $, describing dissipative effects,  is a constant, and $m\vec{f}%
^{(s)}$ is the stochastic force. 

We call Eq. (\ref{eq2}) {\it the
Lorentz-Langevin} equation. As for the random acceleration vector $\vec{f}^{(s)}(t)$ we assume that it is given by the white noise form, with the properties \cite{Bal2,Bal3,Coff}
\bea
\left \langle f^{(s)}_i (t)\right \rangle = 0, \left \langle f^{(s)}_i
\left(t_1\right) f^{(s)}_j \left(t_2\right) \right \rangle &=& \frac{{ \mathcal{A}}}{m^2}\delta _{ij}
\delta \left(t_1 - t_2\right) , \nonumber\\
&&i,j=x,y,z,
\eea
where $\mathcal{A}$ is a normalization constant that can be interpreted as the variance of the random process. 

Hence, in the present approach we assume that the random effects in the Lorentz-Langevin equation can be described by Gaussian
processes, which implies that a complete statistical description of these processes can be obtained from the first and the second order correlation
functions. 

In Eq.~(\ref{eq2}) $\gamma $ can be interpreted physically as an effective collisions frequency, while $-\gamma \vec{v}(t)$ is the
damping term, which describes the average effect of the interparticle collisions. The collision
frequency $\gamma $ and the value $\mathcal{A}$ of the normalization constant are related to the equilibrium thermal velocity $v_{th}$ by the relation $v_{th}^{2}=\left( \mathcal{A}/2\right) \gamma $ \citep{Bal2, Bal3,Coff}, which is also valid for charged particles in a magnetic field.

A current loop in the horizontal $xy$%
-plane, flowing counterclockwise with current intensity $I$, has a dipole
momentum  $\boldsymbol{\mu}_z=\mu \vec{e}_z$, where $\mu =I\times\left(\it{area \;of \;the \;loop}\right)=I\times S$, and $\vec{e}_{z}$ is the unit vector of the $z$
axis. Consequently, the current loop generates a magnetic dipole field, with the dipole momentum
oriented in the positive direction of the $z$-axis. The dipole field can be
derived from the vector potential \citep{Dil}
\begin{eqnarray}  \label{eq3}
\vec{A}&=&\frac{1}{4\pi \varepsilon _{0}c^{2}}\frac{1}{r^{2}}\boldsymbol{\mu}%
_{z}\times \vec{e}_{r}=\frac{1}{4\pi \varepsilon _{0}c^{2}}\frac{1}{r^{3}}%
\boldsymbol{\mu}_{z}\times \vec{r}  \notag \\
&=&M_{z}\frac{1}{r^{3}}\left( -y\vec{e}_{x}+x\vec{e}_{y}\right) ,
\end{eqnarray}
where $r=\sqrt{x^{2}+y^{2}+z^{2}}$, and $M_{z}=\mu /4\pi \varepsilon
_{0}c^{2}$ is the scalar dipole momentum. The vector potential is
independent of the $z$ coordinate.

As $\vec{B}=\nabla \times \vec{A}$, by using Eq. (\ref{eq3}), the
Lorentz-Langevin equations Eq. (\ref{eq2}) describing the motion of a charge
particle in a dipole field in the presence of dissipation and of a
stochastic force are given by
\begin{equation}
\frac{d^{2}x}{dt^{2}}=3\alpha \frac{z}{r^{5}}\left( \dot{y}z-\dot{z}y\right)
-\alpha \frac{1}{r^{3}}\dot{y}-\gamma \dot{x}+f_{x}^{(s)},  \label{LL1}
\end{equation}
\begin{equation}
\frac{d^{2}y}{dt^{2}}=-3\alpha \frac{z}{r^{5}}\left( \dot{x}z-\dot{z}%
x\right) +\alpha \frac{1}{r^{3}}\dot{x}-\gamma \dot{y}+f_{y}^{(s)},
\label{LL2}
\end{equation}
\begin{equation}
\frac{d^{2}z}{dt^{2}}=3\alpha \frac{z}{r^{5}}\left( \dot{x}y-\dot{y}x\right)
-\gamma \dot{z}+f_{z}^{(s)},  \label{LL3}
\end{equation}%
where we have denoted $\alpha =qM_{z}/m$. In the absence of the damping and of the stochastic forces, with $\gamma =0$, and $f_i^{(s)}=0$, $i=x,y,z$, the system of Eqs.~(\ref{LL1})-(\ref{LL3} reduces to the classical St\"{o}rmer problem in Cartesian coordinates \citep{Dil}.  

Obtaining the solutions of Eqs.~(\ref{LL1})-(\ref{LL3}) in the presence of
dissipation and stochastic forces represents {\it the Stochastic-Dissipative St\"{o}rmer
Problem (SDSP)}, already mentioned in the Introduction Section.

\subsection{Dimensionless form of the Stochastic-Dissipative St\"{o}rmer Problem (SDSP) evolution equations}

We now rescale the system of Eqs. (\ref{LL1})-(\ref{LL3}) by introducing a
set of dimensionless quantities
\begin{equation}
X=\frac{x}{r_{0}},Y=\frac{y}{r_{0}},Z=\frac{z}{r_{0}},
\end{equation}%
where $r_{0}$ is a specific length, like, for example, the radius of the
Earth $r_{0}=6378136$ m. Then the system of Eqs. (\ref{LL1})-(\ref{LL3})
becomes
\begin{equation}
\frac{d^{2}X}{dt^{2}}=3\beta \frac{Z}{R^{5}}\left( \dot{Y}Z-\dot{Z}Y\right)
-\beta \frac{1}{R^{3}}\dot{Y}-\gamma \dot{X}+F_{x}^{(s)},
\end{equation}
\begin{equation}
\frac{d^{2}Y}{dt^{2}}=-3\beta \frac{Z}{R^{5}}\left( \dot{X}Z-\dot{Z}X\right)
+\beta \frac{1}{R^{3}}\dot{X}-\gamma \dot{Y}+F_{y}^{(s)},
\end{equation}
\begin{equation}
\frac{d^{2}Z}{dt^{2}}=3\beta \frac{Z}{R^{5}}\left( \dot{X}Y-\dot{Y}X\right)
-\gamma \dot{Z}+F_{z}^{(s)},
\end{equation}%
where $\beta =\alpha /r_{0}^{3}$, $R=\sqrt{X^{2}+Y^{2}+Z^{2}}$, and $%
F_{x}^{(s)}=f_{x}^{(s)}/r_{0}$ etc.

We finally rescale the time coordinate
according to $\tau =\beta t$, thus obtaining the dimensionless form of the
Lorentz-Langevin system for the dipole magnetic field as

\begin{equation}
\frac{d^{2}X}{d\tau ^{2}}=3\frac{Z}{R^{5}}
\left( \frac{dY}{d\tau }Z-\frac{dZ}{d\tau }Y\right) -\frac{1}{R^{3}}\frac{dY%
}{d\tau }-\Gamma \frac{dX}{d\tau }+\Phi _{x}^{(s)},  \label{NN1}
\end{equation}%
\be
\frac{d^{2}Y}{d\tau ^{2}}=-3\frac{Z}{R^{5}}
\left( \frac{dX}{d\tau }Z-\frac{dZ}{d\tau }X\right) +\frac{1}{R^{3}}\frac{dX%
}{d\tau }-\Gamma \frac{dY}{d\tau }+\Phi _{y}^{(s)},  \label{NN2}
\ee

\begin{equation}
\frac{d^{2}Z}{d\tau ^{2}}=3\frac{Z}{R^{5}}\left( \frac{dX}{d\tau }Y-\frac{dY%
}{d\tau }X\right) -\Gamma \frac{dZ}{d\tau }+\Phi _{z}^{(s)},  \label{NN3}
\end{equation}%
where $\Gamma =\gamma /\beta $ and $\Phi _{x}^{(s)}=$ $F_{x}^{(s)}/\beta ^{2}
$ etc. When $R(\tau )=1$, the particle reaches the surface of the body
creating the dipole field, and hence the initial conditions must be chosen
so that $R(\tau )>1$.

We also introduce the dimensionless velocity of the particle $\vec{V}$, defined according to
\begin{equation}
\vec{V}=\left(V_x,V_y,V_z\right)=\left(\frac{dX}{d\tau},\frac{dY}{d\tau},\frac{dZ}{d\tau}\right).
\end{equation}

\subsection{Energy losses, and radiation}

In the nonrelativistic limit the total electromagnetic power $P$ emitted by a moving charge is given by  \citep{LaLi},
\begin{equation}
P=\frac{q^2}{6\pi \epsilon _0c^3}\vec{a}^{2},
\end{equation}%
where $\vec{a}=d\vec{v}/dt$ is the acceleration of the particle.

By taking into account that for a charged particle moving in a magnetic field in the presence of dissipative and stochastic forces
the acceleration is given by Eq.~(\ref{eq2}), for the total electromagnetic power
emitted by the randomly moving particle we obtain the expression
\bea
P=\frac{q^2}{6\pi \epsilon _0c^3}\left[\frac{q}{m}\vec{v}\times \vec{B}-\gamma \vec{v}+\vec{f}^{(s)}\right]^2.
\eea
In a dimensionless form the power is given by
\bea
P&=&\frac{q^2r_0\beta ^2}{6\pi \epsilon _0c^3}\left[\left(\frac{d^2X}{d\tau ^2}\right)^2+\left(\frac{d^2Y}{d\tau ^2}\right)^2+\left(\frac{d^2Z}{d\tau ^2}\right)^2\right]\nonumber\\
&=&\frac{q^2r_0\beta ^2}{6\pi \epsilon _0c^3}\tilde{P},
\eea
where
\be
\tilde{P}=\left(\frac{d^2X}{d\tau ^2}\right)^2+\left(\frac{d^2Y}{d\tau ^2}\right)^2+\left(\frac{d^2Z}{d\tau ^2}\right)^2.
\ee

 We define the average kinetic energy $K$ of the system according to
 \be
 K=\frac{m}{2}\left<\vec{v}^2\right>.
 \ee

 By multiplying Eq.~(\ref{eq2}) by $\vec{v}$, and taking the average we obtain
 \be\label{enbal}
 \frac{dK}{dt}=-2\gamma K+\left<\vec{v}\cdot \vec{f}^s\right>.
 \ee

 From a physical point of view we can interpret the term $-2\gamma K$ as corresponding to the energy dissipation, while $W=\left<\vec{v}\cdot \vec{f}^s\right>$ is the work done on the system by the external forces \citep{Huang}. The work $W$ can be obtained as
\begin{equation}
\left\langle \vec{v}(t)\cdot \vec{f}^{(s)}\left(t'\right)\right\rangle =\left\{
\begin{array}{c}
\frac{\mathcal{A}}{m^{3}}e^{-\gamma \left( t-t^{\prime }\right) },t>t^{\prime }, \\
0,\;\;\;\;\;\;\;\;\;\;\;\;\;\;\;\;\;t<t^{\prime },%
\end{array}%
\right. .
\end{equation}

In the limit $t\rightarrow t'$ we obtain for the average value of the work done on the particle the expression
\be
\left\langle \vec{v}\cdot \vec{f}^{(s)}\left(t'\right)\right\rangle=\frac{\mathcal{A}}{2m^3}.
\ee

Then the energy balance equation (\ref{enbal}) takes the form
\be
\frac{dK}{dt}=\frac{\mathcal{A}}{2m^3}-2\gamma K,
\ee
\textcolor{red}{and can be integrated to give}
\be
K=\frac{\left(4 \gamma  K_0 m^3-\mathcal{A}\right)}{4 \gamma  m^3}e^{-2 \gamma  t}+\frac{\mathcal{A}}{4
   \gamma  m^3},
\ee
where $K_0=K(0)$ is the initial kinetic energy of the particles. 
In the asymptotic limit $t\rightarrow \infty$, we obtain $K\rightarrow \mathcal{A}/2m^3$, that is, the average value of the kinetic energy becomes a constant. For $\gamma =0$, that is, in the absence of dissipation, we obtain
\be
K(t)=\frac{\mathcal{A}}{2m^3}t+K_0,
\ee
where $K_0=K(0)$ is an integration constant, giving again the initial kinetic energy. Hence, under the action of the stochastic forces only, the kinetic energy of the particle increases linearly in time.

\subsection{The numerical scheme}

Eqs.~(\ref{NN1})-(\ref{NN3}) can be generally solved only numerically. To obtain their solution we use a multidimensional Milstein scheme \citep{Mi}. The equations of motion can be rewritten in an update form as
\begin{equation}
d\vec{S}(\tau) = \vec{C}(\vec{S})d\tau + \bar{D}(\vec{S})d\vec{W}(\tau),
\end{equation}
where
\begin{equation}
d\vec{S}(\tau) = (dV_x, dX, dV_y, dY, dV_z, dZ)^T,
\end{equation}
$\vec{C}(\vec{S})$ is a six dimensional vector with components
\begin{equation}
C_1(\vec{S}) = 3\frac{Z}{R^{5}}
\left( V_yZ-V_zY\right) -\frac{1}{R^{3}}V_y-\Gamma V_x,
\ee
\be
C_2(\vec{S}) = V_x,
\end{equation}
\begin{equation}
C_3(\vec{S}) = -3\frac{Z}{R^{5}}
\left( V_xZ-V_zX\right) +\frac{1}{R^{3}}V_x-\Gamma V_y ,
\ee
\be
C_4(\vec{S}) = V_y,
\end{equation}
\begin{equation}
C_5(\vec{S}) = 3\frac{Z}{R^{5}}\left( V_xY-V_yX\right) -\Gamma V_z,
\ee
\be
C_6(\vec{S}) = V_z,
\end{equation}
and $\bar{D}(\vec{S})$ is a $6\times 6$ matrix with the only nonzero components 
\be
\bar{D}_{11}=\bar{D}_{33}=\bar{D}_{55}=1.
\ee

The noise term is a six dimensional vector
\begin{equation}
d\vec{W}(\tau)=(dW_x(\tau),0,dW_y(\tau),0,dW_z(\tau)),
\end{equation}
respectively. $dW_i$ is a zero mean Wienner process.

For the numerical implementation the solution vector is discretised in units $h$ of the independent variable $\tau$ as $\tau = nh$, such that the solution is advanced as
\begin{equation}
S_i(n+1)=S_i(n)+C_i\left (\vec{S}(n)\right)h+\bar{D}_{ij}(n)dW_i(n)\delta _{ij}.
\end{equation}

Note that the Milstein scheme usually contains another term, which for this application cancels due to the simple form of the matrix $\bar{D}(\vec{S})$. This scheme was thus implemented specifically for the Stochastic Dissipative St\"{o}rmer Problem (SDSP) as
\begin{equation}
S_i(n+1)=S_i(n)+C_i\left (\vec{S}(n)\right)h+\sqrt{\sigma _{\Phi i} ^2 h} N_{1i}(n),
\end{equation}
where the number $N_{1i}(n)$ is drawn for each timestep and for each $S_i$ from a standard unit normal.

To obtain the radiation power, one needs to compute the accelerations. Since the matrix $\bar{D}(\vec{S})$ has a simple form, the equation for the accelerations becomes
\begin{equation}
a_i = C_i(\vec{S})+\sigma _{\Phi i} N_{2i},i=1,2,3,
\end{equation}
where the number $N_{2i}(n)$ is drawn for each timestep and for each $a_i$ from a standard unit normal.
The corresponding noise (being the formal derivative of the Wiener process) is drawn from a distribution $\mathcal{N}(0, \sigma _G=\sigma _W/ \sqrt{h} )$. We present the radiation patterns of the above trajectories in terms of $\sigma_W$, keeping in mind the connection between the two volatilities.

\section{Trajectories and radiation patterns in the Stochastic-Dissipative St\"{o}rmer Problem (SDSP)}\label{sect2}

In the present Section we perform a detailed numerical investigation of the Stochastic-Dissipative St\"{o}rmer problem. We obtain both the particle trajectories, as well as the power radiated by the particle in motion in the magnetic dipolar field. The radiation emitted by such a particle in motion is analyzed via its power spectral density (PSD). 
The slopes and the numerical values of the PSD function could give some important insights into the nature of the variability observed in the physical processes.  

Let's assume that $X$ is a stationary fluctuating quantity, having the mean $\mu _X$ and the variance $\sigma _X ^2$. The autocorrelation function for  $X$ is defined according to~\citep{Vaseghi,Larsen}
\begin{equation}
R_{XX}(\tau) = \frac{\langle \left ( X_s - \mu _X \right ) \left (
X_{s+\tau} - \mu \right) \rangle}{\sigma _X^2},
\end{equation}
where by $X_s$ we have denoted  the values of $X$ measured at the time $s$. Moreover, by $\langle
\rangle$ we have denoted the averaging over all values $s$. 

The PSD associated to a stochastic variable is defined with the use of the correlation function as~\cite{Vaseghi,Larsen}
\begin{equation}
PSD(\omega ) = \int _{-\infty} ^{+\infty} R_{XX}(\tau) e^{-\imath 2\pi \omega  \tau}
d\tau. 
\end{equation}
Hence, the PSD of a signal is the Fourier transform of the autocorrelation function of that signal.
The importance of the PSD can be understood in terms of the
"memory" of the considered process. The slope of the PSD of a time series of
$X$ give important insights into the degree of correlation the considered
physical process has with itself. 

A stochastic process of Brownian type has a PSD of the form $PSD(\omega) \sim \omega ^{-2}$.  A completely uncorrelated evolution of a physical system is characterized by  white noise, with the PSD having the shape $PSD(\omega) = \omega ^0 = {\rm const}$. Hence, the slope of the PSD is an indicator of the type of physical process generating a stochastic signal.


We begin our analysis with the trajectories and radiation patterns in the Classical St\"{o}rmer Problem (CSP), and then we are going to investigate the effects of dissipation and stochastic effects on the dynamical evolution. All results are for $h=0.001$.

\subsection{Classical St\"{o}rmer Problem-CSP (no friction and no Brownian Motion)}

We begin, for the sake of comparison, our analysis with the classical St\"{o}rmer problem, without dissipation and stochastic effects. In the $Z=0$ plane the equations of motion of a nonrelativistic  charged particle in the dipole magnetic field are
\begin{equation}\label{CSP1}
\frac{d^{2}X}{d\tau ^{2}}= -\frac{1}{R^{3}}\frac{dY%
}{d\tau },
\end{equation}%
\be\label{CSP2}
\frac{d^{2}Y}{d\tau ^{2}}= \frac{1}{R^{3}}\frac{dX%
}{d\tau }.
\ee

After multiplying Eq.~(\ref{CSP1}) by $dX/d\tau$, Eq.~(\ref{CSP2}) by $dY/d\tau$, and adding the two equations, it follows that the system of equations (\ref{CSP1}) and (\ref{CSP2}) admits the first integral
\be
V^2=\left(\frac{dX}{d\tau}\right)^2+\left(\frac{dY}{d\tau}\right)^2=C^2,
\ee
where $C^2$ is an integration constant. This equation expresses the conservation of the kinetic energy of the particle. Equations.~(\ref{CSP1}) and (\ref{CSP2}) admit a particular solution of the form
\begin{equation}\label{ss}
X\left(\tau \right) =\frac{1}{\theta ^{1/3}}\sin \left(\theta \tau +\phi\right)
,Y\left(\tau\right) =-\frac{1}{\theta ^{1/3}}\cos \left( \theta \tau +\phi\right),
\end{equation}
where $\theta >0$ and $\phi $ are constants, which are determined from the initial conditions as $\tan \phi=-X_0/Y_0$, and $\theta =1/\left(X_0^2+Y_0^2\right)^{3/2}$, respectively.

\subsubsection{Numerical results}

In our numerical simulations the particle is injected into the dipole field from an initial position $\vec{R}_0=\left(X_0,Y_0,Z_0\right)$, with initial velocities $\vec{V}_0=\left(V_{x0},V_{y0}, V_{z0}\right)$.  

The solution for two CSP cases was obtained and it was verified by the $K1-0$ method of \cite{gott2004} that one case is not chaotic, and one is chaotic. The solutions differ only by their initial conditions and the purpose is to show how the behavior of these two different classes of solutions changes as we add more complexity to the problem.

\begin{figure*}[htb!]
\centering
\includegraphics[width = 0.31\textwidth]{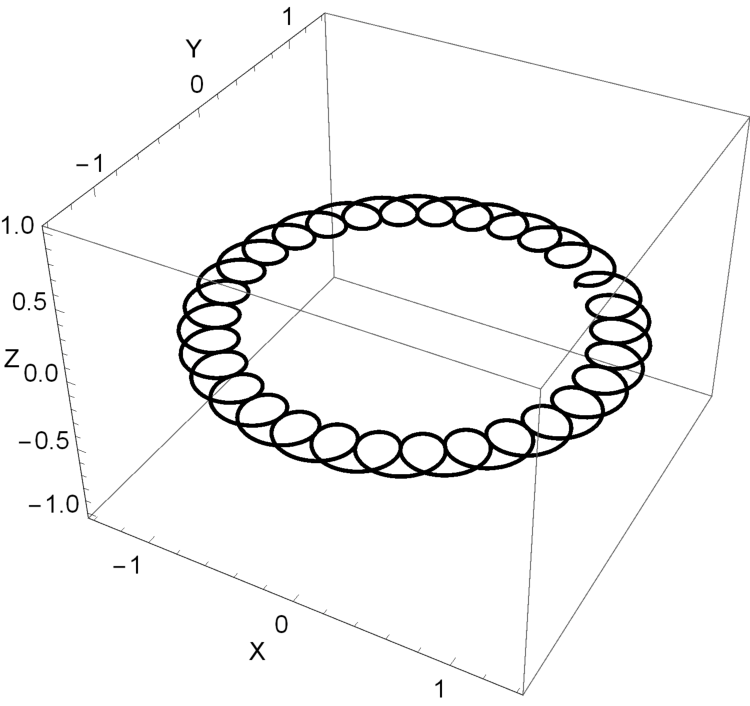}
\includegraphics[width = 0.31\textwidth]{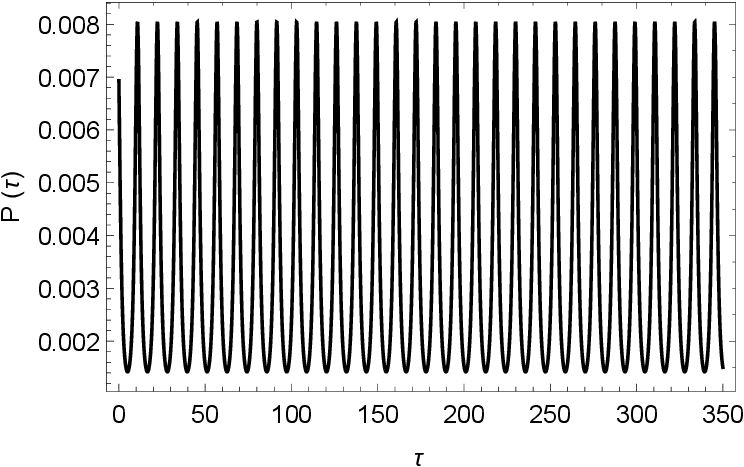}
\includegraphics[width = 0.31\textwidth]{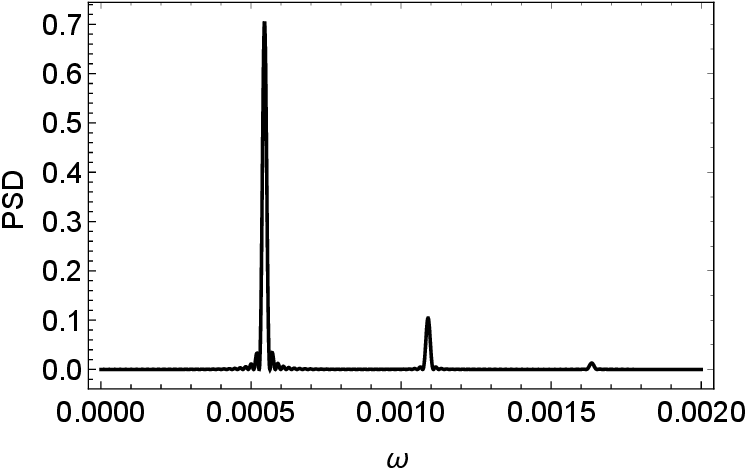}
\caption{Periodic motion, radiation power, and PSD in the Classical St\"{o}rmer Problem (CSP) for $\vec{R}_0=\left(0.7,0.8,0\right)$, $\left|\vec{R}_0\right|=1.063$, and $\vec{V}_0=\left(0.10,0,0\right)$. For the numerical simulations the values  $h=0.001$ and $L=350000$ have been adopted.}\label{fig:CSP1}
\end{figure*}

\begin{figure*}[htb!]
\centering
\includegraphics[width = 0.31\textwidth]{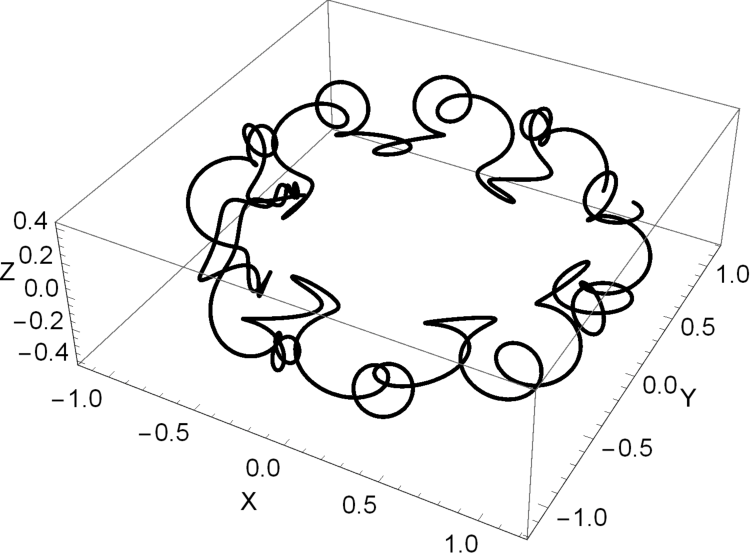}
\includegraphics[width = 0.31\textwidth]{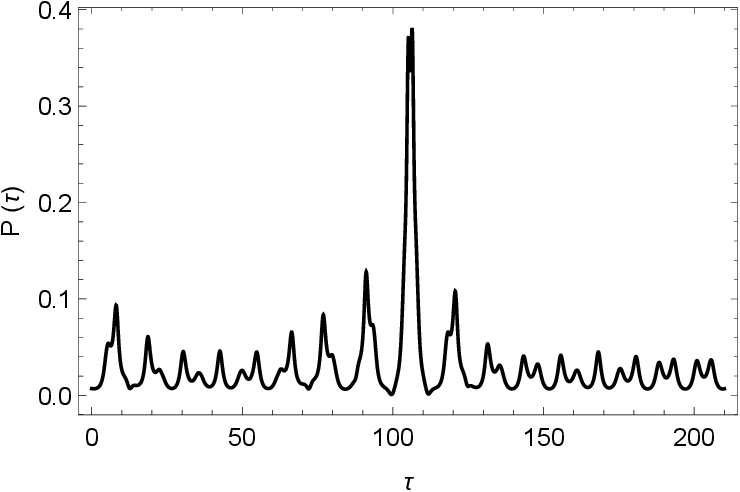}
\includegraphics[width = 0.31\textwidth]{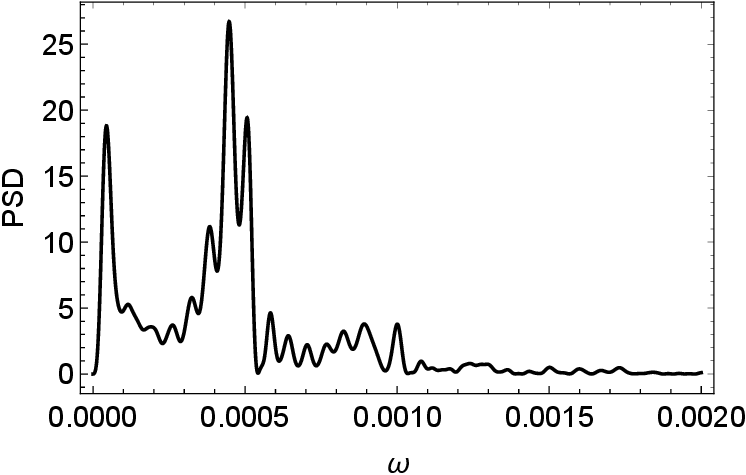}
\caption{Chaotic motion, radiation power and PSD in the Classical St\"{o}rmer Problem (CSP) for $\vec{R}_0=\left(0.7,0.8,0\right)$, $\left|\vec{R}_0\right|=1.063$, and $\vec{V}_0=\left(0.01,0.10,0.10\right)$. For the numerical simulations the values  $h=0.001$ and $L=180000$ have been adopted.}\label{fig:CSP4}
\end{figure*}

Figs.~\ref{fig:CSP1}- \ref{fig:CSP4} show the three dimensional trajectory, radiation power and PSD of the radiation for a particle in the CSP for the periodic and chaotic trajectories. As one can see from Fig.~\ref{fig:CSP1}, describing a strictly periodic motion, the emitted electromagnetic power has also a periodic structure,  with constant maximum values of $P$. The PSD is generally constant, however, two peaks can be observed in its structure. 

For the chaotic case of the CSP problem, as illustrated in Fig.~\ref{fig:CSP4}, the radiation spectrum shows a significant difference as compared to the strictly periodic case, with a sharp maximum in the emitted power, corresponding to a stochastic resonance type phenomenon. The spectrum of the radiation also indicates the presence of stochastic characteristics.    

\subsection{Classical Dissipative St\"{o}rmer Problem - CDSP (friction, and no Brownian Motion)}

In the presence of dissipation, described in terms of a friction force,  the dimensionless equations of motion in the $Z=0$ plane of a particle in a dipole magnetic field are given by
\begin{equation}\label{DSP1}
\frac{d^{2}X}{d\tau ^{2}}= -\frac{1}{R^{3}}\frac{dY%
}{d\tau }-\Gamma \frac{dX}{d\tau},
\end{equation}%
\be\label{DSP2}
\frac{d^{2}Y}{d\tau ^{2}}= \frac{1}{R^{3}}\frac{dX%
}{d\tau }-\frac{dY}{d\tau}.
\ee

After multiplying Eq.~(\ref{DSP1}) by $dX/d\tau$, Eq.~(\ref{DSP2}) by $dY/d\tau$, and adding the resulting equations we obtain
\be
\frac{d}{d\tau}\left[\left(\frac{dX}{d\tau}\right)^2+\left(\frac{dY}{d\tau}\right)^2\right]=-2\Gamma \left[\left(\frac{dX}{d\tau}\right)^2+\left(\frac{dY}{d\tau}\right)^2\right].
\ee
Integrating once the above equation we find the first integral of the Dissipative St\"{o}rmer Problem as given by
\be
V^2=\left(\frac{dX}{d\tau}\right)^2+\left(\frac{dY}{d\tau}\right)^2=C^2e^{-2\Gamma \tau},
\ee 
where $C^2$ is an integration constant. For $\Gamma =0$ we recover the first integral of the Classical St\"{o}rmer Problem. In the limit $\tau \rightarrow \infty$, we have $\lim _{\tau \rightarrow \infty}V^2=0$, indicating a decrease in the particle velocity due to energy loss.  

\begin{figure*}[htb!]
\centering
\includegraphics[width = 6.5cm, height=4.25cm]{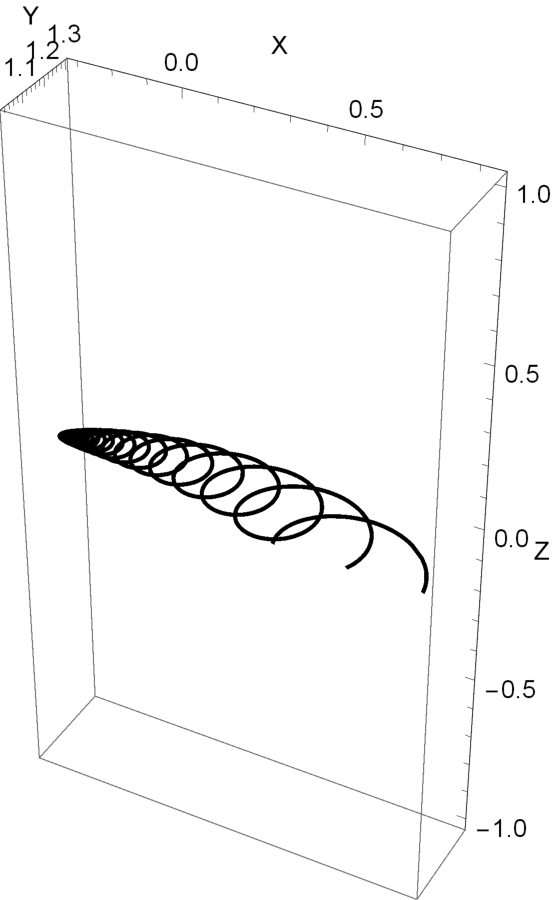}
\includegraphics[width = 0.31\textwidth]{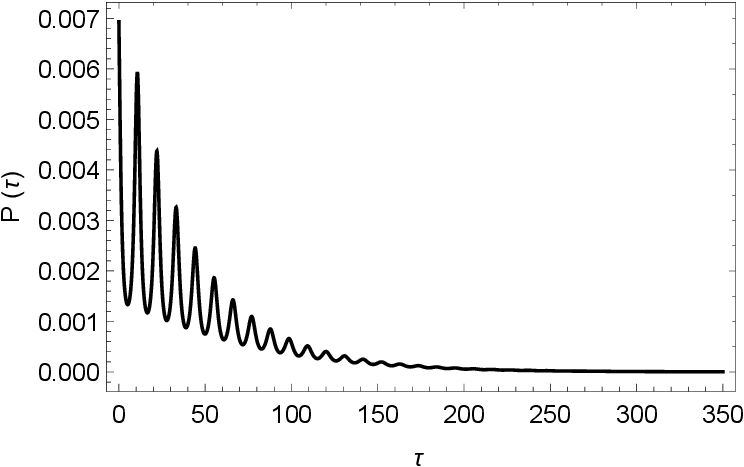}
\includegraphics[width = 0.31\textwidth]{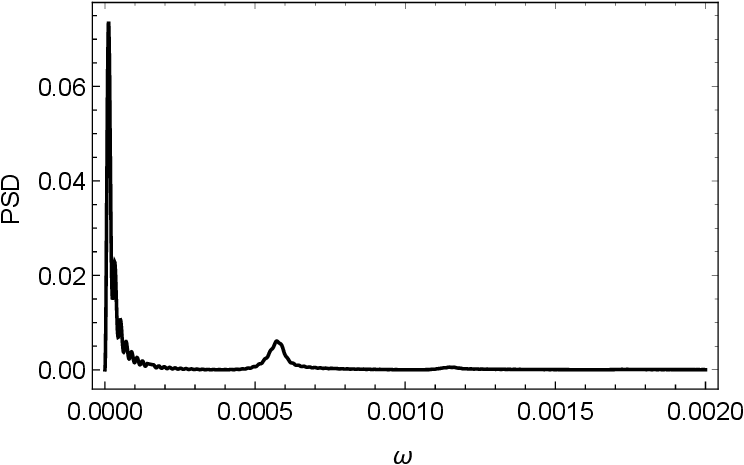}
\caption{Periodic motion in the Classical Dissipative St\"{o}rmer Problem (CDSP): trajectory, radiation and PSD for $\vec{R}_0=\left(0.7,0.8,0\right)$, $\left|\vec{R}_0\right|=1.063$, and $\vec{V}_0=\left(0.10,0,0\right)$, $h=0.001$, $L=350000$ and $\Gamma=10^{-2}$, respectively. Due to increasing friction, the particle no longer covers the $xOy$ plane as in the CSP.}\label{fig:CSP1-f1e-2}
\end{figure*}

Figure~\ref{fig:CSP1-f1e-2} shows the effect of nonzero friction on the CSP periodic trajectory, radiation and PSD of radiation. The presence of friction drastically changes the patterns of motion, radiation emission, and the PSD, with the electromagnetic radiation emission rapidly tending to zero, and with a rapid decrease of the peaks of the radiation maximum. The PSD also tends rapidly to a constant value. 

\begin{figure*}[htb!]
\centering
\includegraphics[width = 0.31\textwidth]{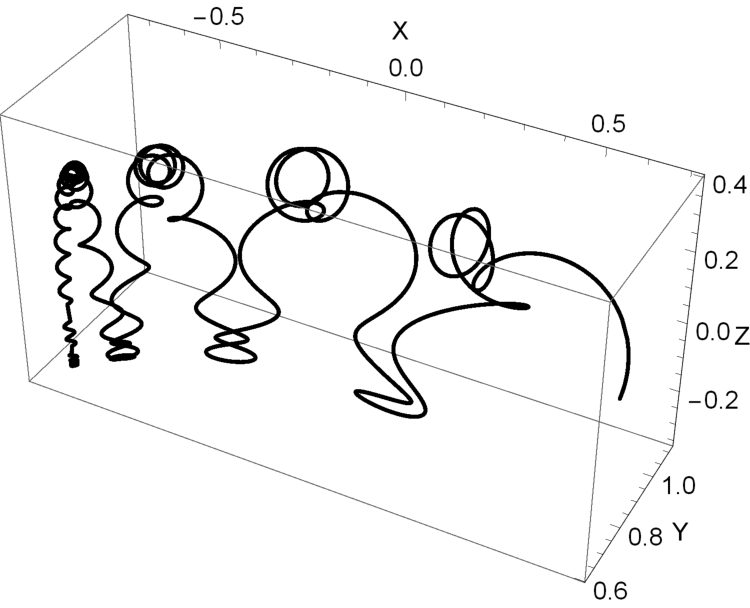}
\includegraphics[width = 0.31\textwidth]{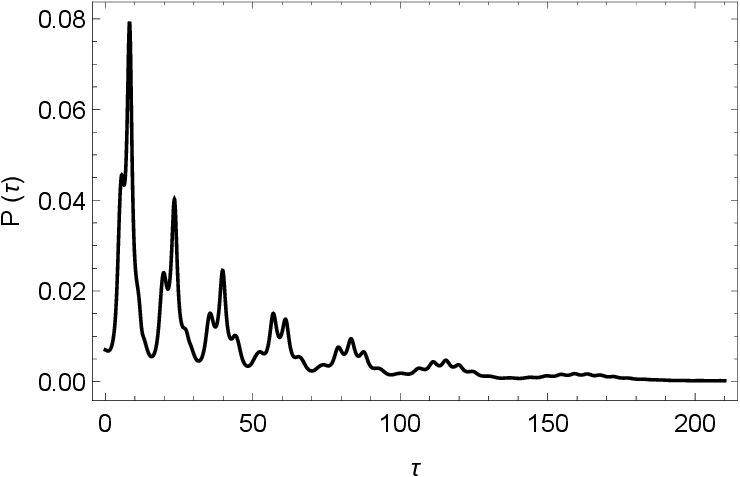}
\includegraphics[width = 0.31\textwidth]{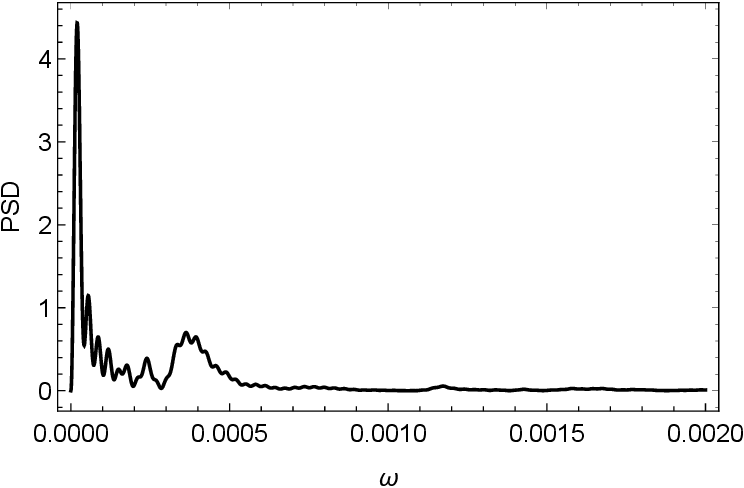}
\caption{Chaotic motion in the Classical Dissipative St\"{o}rmer Problem (CDSP): trajectory, radiation and PSD for $\vec{R}_0=\left(0.7,0.8,0\right)$, $\left|\vec{R}_0\right|=1.063$, and $\vec{V}_0=\left(0.01,0.10,0.10\right)$, $h=0.001$, $L=180000$ and $\Gamma=10^{-2}$.}\label{fig:CSP4-f1e-2}
\end{figure*}

Figure~\ref{fig:CSP4-f1e-2} describes the effects of increasing friction on the CSP chaotic trajectories, radiation and PSD of radiation. The radiation power tends to zero in the large time limit, with a series of distinct and well defined peaks in the electromagnetic power emission. The PSD also tends to zero, with a similar behavior as in the periodic dissipative case.   

\subsection{Brownian motion in the Stochastic-Dissipative St\"{o}rmer Problem - SDSP-dissipation and stochastic effects}

\begin{figure*}[htb!]
\centering
\includegraphics[width = 0.31\textwidth]{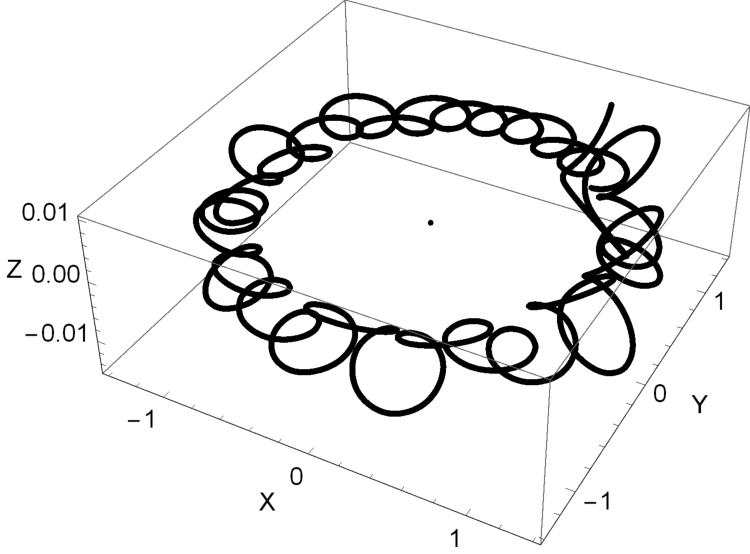}
\includegraphics[width = 0.31\textwidth]{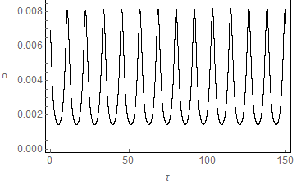}
\includegraphics[width = 0.31\textwidth]{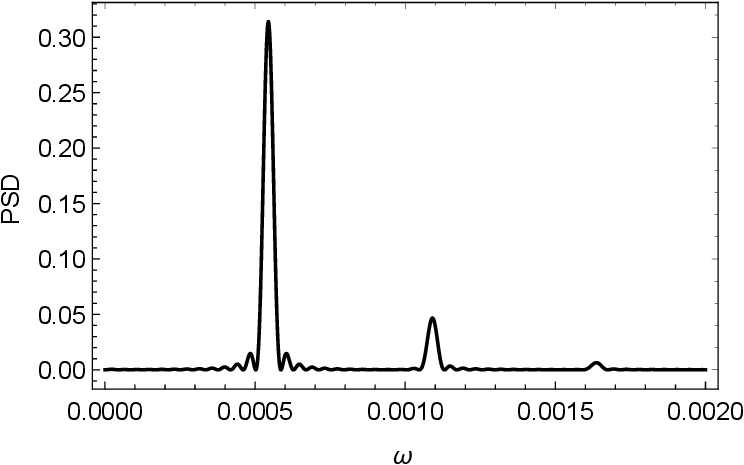}
\caption{Periodic motion in  the Stochastic-Dissipative St\"{o}rmer Problem (SDSP): trajectory, radiation and PSD for $\vec{R}_0=\left(0.7,0.8,0\right)$, $\left|\vec{R}_0\right|=1.063$, and $\vec{V}_0=\left(0.10,0,0\right)$, $h=0.001$, $L=150000$, for $\sigma _S= 10^{-6}$ and $\Gamma = 10^{-4} $.}\label{fig5}
\end{figure*}

\begin{figure*}[htb!]
\centering
\includegraphics[width = 0.31\textwidth]{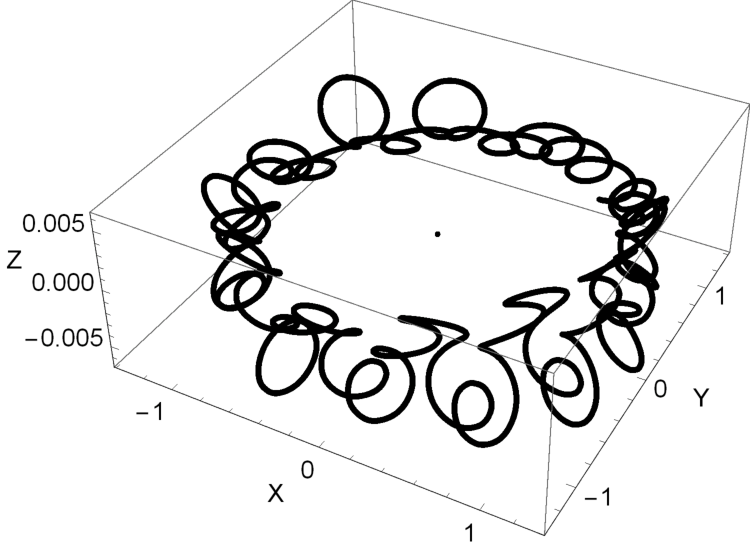}
\includegraphics[width = 0.31\textwidth]{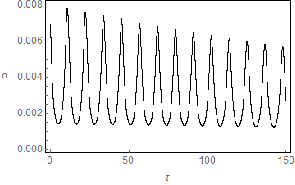}
\includegraphics[width = 0.31\textwidth]{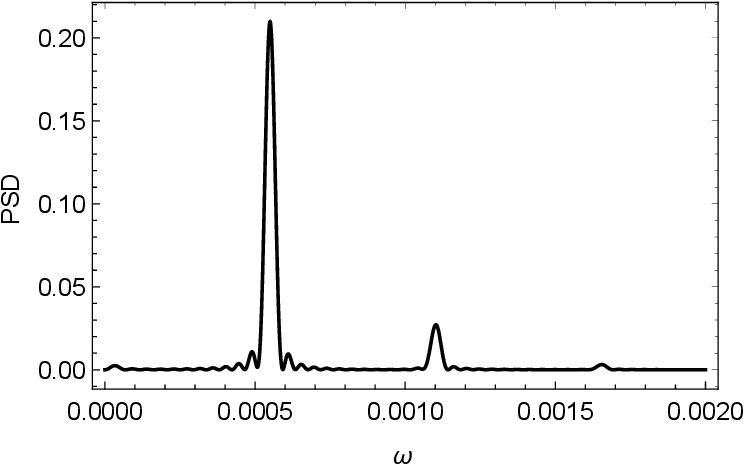}
\caption{Periodic motion in the Stochastic-Dissipative St\"{o}rmer Problem  (SDSP):  trajectory, radiation and PSD for $\vec{R}_0=\left(0.7,0.8,0\right)$, $\left|\vec{R}_0\right|=1.063$, and $\vec{V}_0=\left(0.10,0,0\right)$, $h=0.001$, $L=150000$, for $\sigma _S= 10^{-6}$ and $\Gamma = 10^{-3} $.}\label{fig6}
\end{figure*}

\begin{figure*}[htb!]
\centering
\includegraphics[width = 0.31\textwidth]{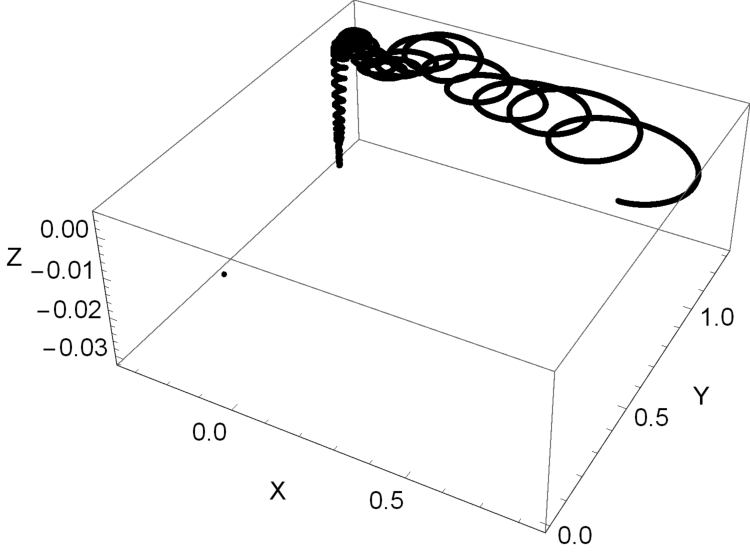}
\includegraphics[width = 0.31\textwidth]{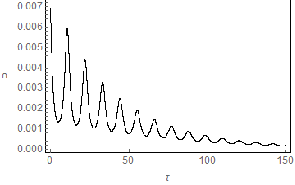}
\includegraphics[width = 0.31\textwidth]{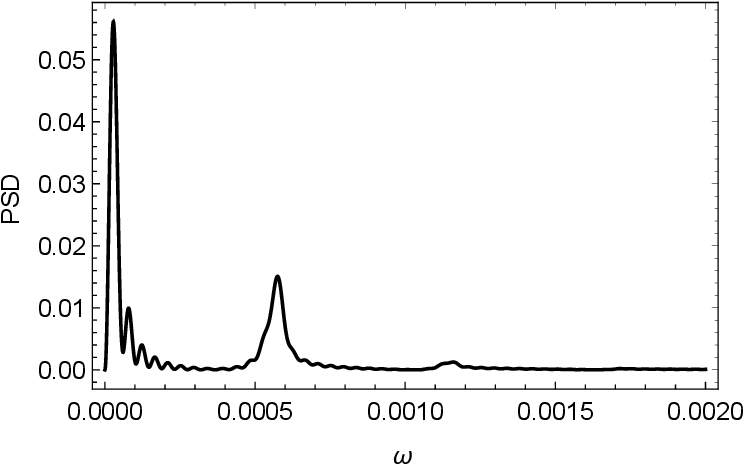}
\caption{Periodic motion in the Stochastic-Dissipative St\"{o}rmer Problem (SDSP):  trajectory, radiation and PSD for $\vec{R}_0=\left(0.7,0.8,0\right)$, $\left|\vec{R}_0\right|=1.063$, and $\vec{V}_0=\left(0.10,0,0\right)$, $h=0.001$, $L=150000$, for $\sigma _S= 10^{-6}$ and $\Gamma = 10^{-2} $.}\label{fig7}
\end{figure*}

\begin{figure*}[htbp!]
\centering
\includegraphics[width = 0.31\textwidth]{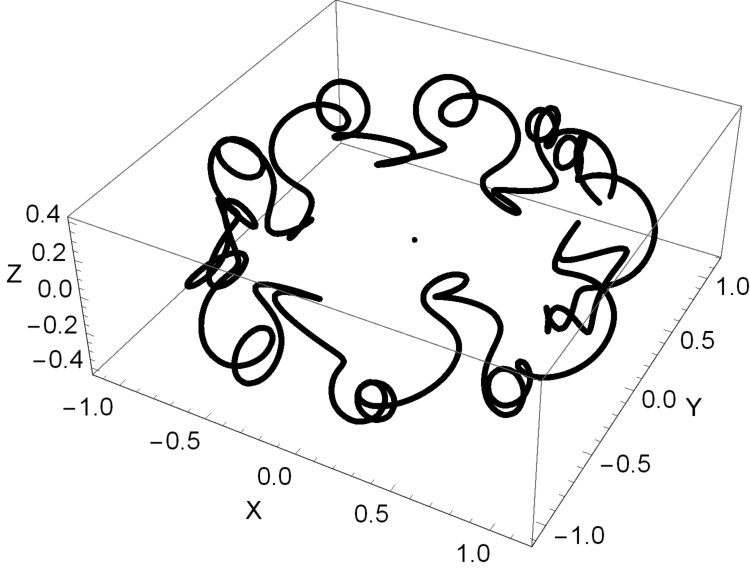}
\includegraphics[width = 0.31\textwidth]{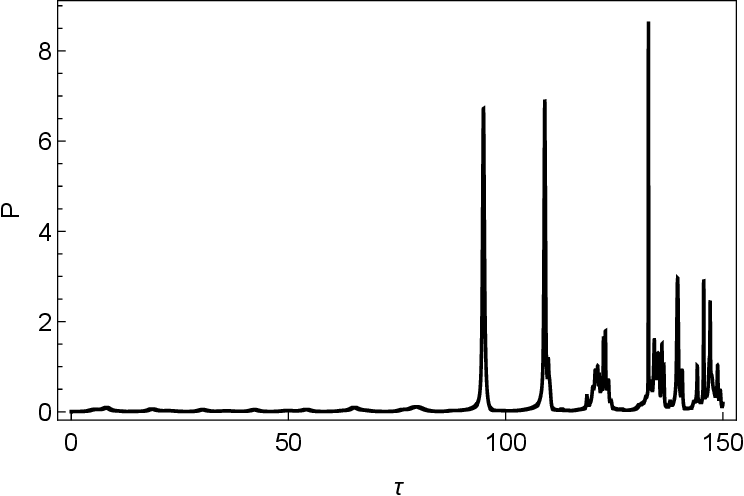}
\includegraphics[width = 0.31\textwidth]{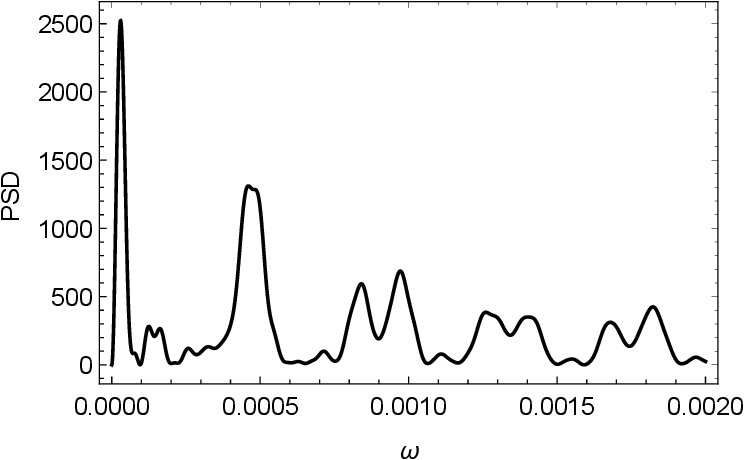}
\caption{Chaotic motion in the Stochastic-Dissipative St\"{o}rmer Problem  (SDSP):  trajectory, radiation and PSD for $\vec{R}_0=\left(0.7,0.8,0\right)$ and $\vec{V}_0=\left(0.01,0.10,0.10\right)$, $L=150000$, for $\sigma _S = 10^{-6}$ and  $\Gamma  = 10^{-4} $.}\label{fig8}
\end{figure*}

\begin{figure*}[htb!]
\centering
\includegraphics[width = 0.31\textwidth]{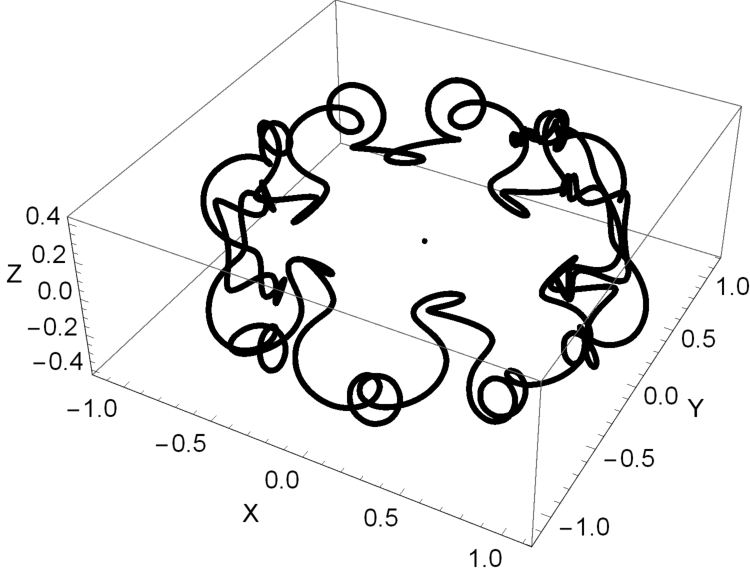}
\includegraphics[width = 0.31\textwidth]{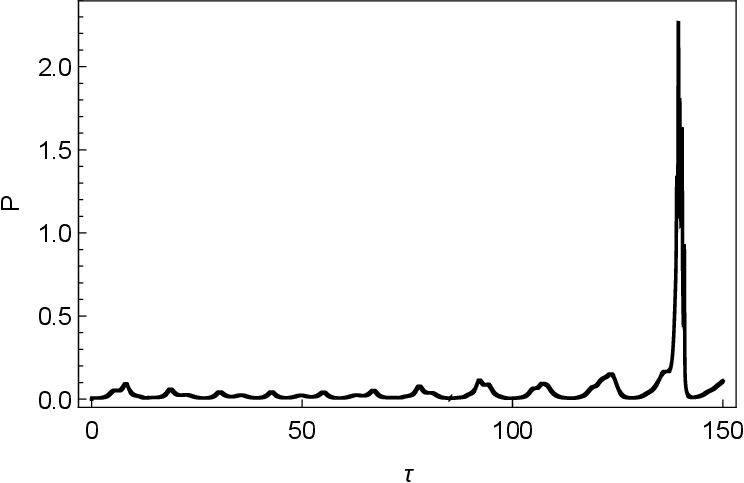}
\includegraphics[width = 0.31\textwidth]{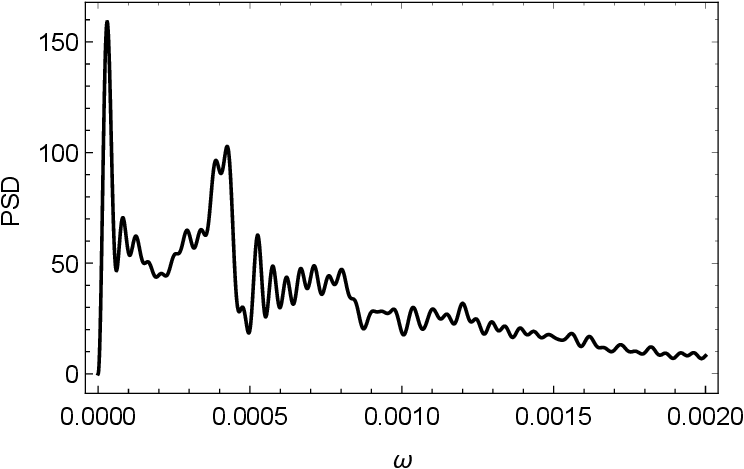}
\caption{Chaotic motion in the Stochastic-Dissipative St\"{o}rmer Problem (SDSP): trajectory, radiation and PSD for $\vec{R}_0=\left(0.7,0.8,0\right)$, $\left|\vec{R}_0\right|=1.063$, and $\vec{V}_0=\left(0.01,0.10,0.10\right)$, $L=150000$, for $\sigma _S = 10^{-6}$ and  $\Gamma  = 10^{-3} $.}\label{fig9}
\end{figure*}

\begin{figure*}[htb!]
\centering
\includegraphics[width = 0.31\textwidth]{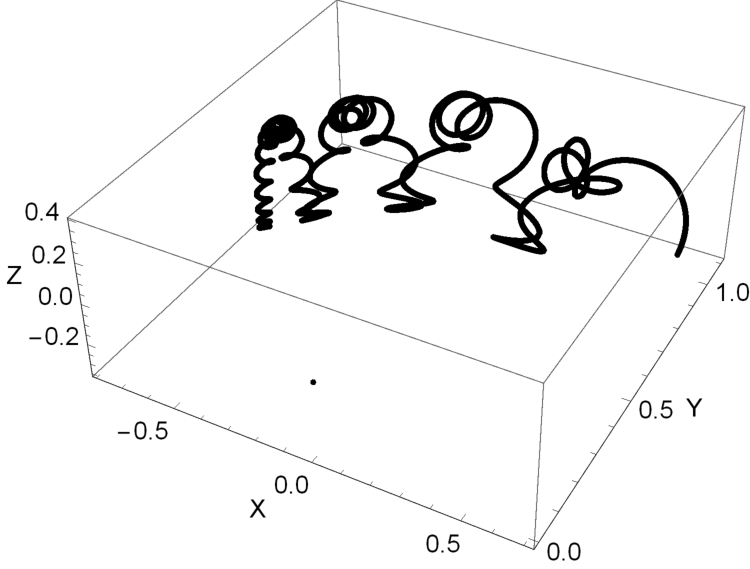}
\includegraphics[width = 0.31\textwidth]{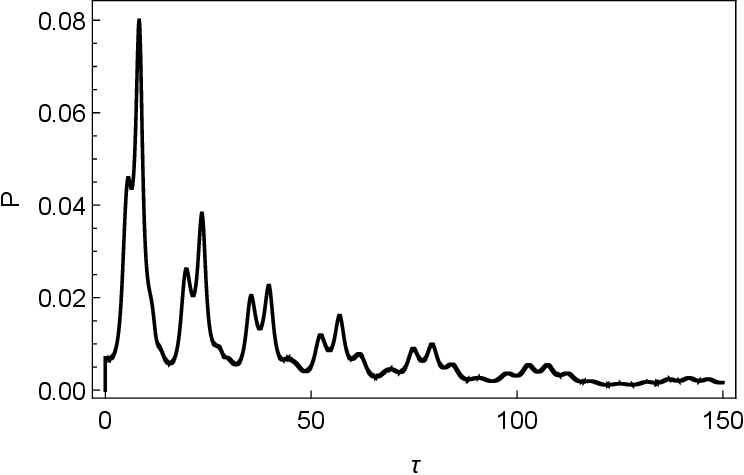}
\includegraphics[width = 0.31\textwidth]{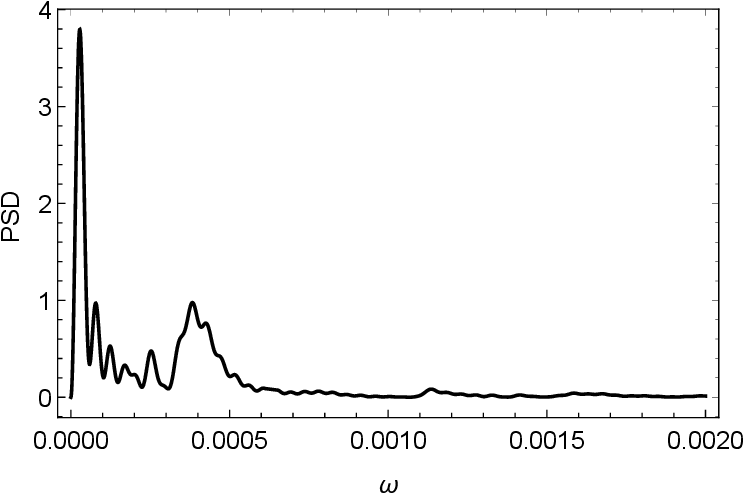}
\caption{Chaotic motion in the Stochastic-Dissipative St\"{o}rmer Problem (SDSP):  trajectory, radiation and PSD for $\vec{R}_0=\left(0.7,0.8,0\right)$, $\left|\vec{R}_0\right|=1.063$, and $\vec{V}_0=\left(0.01,0.10,0.10\right)$, $L=150000$, for $\sigma _S = 10^{-6}$ and  $\Gamma  = 10^{-2} $}\label{fig10}
\end{figure*}

For Brownian motion in the St\"{o}rmer problem, trajectories, radiation patterns and PSD of the charged particles are presented in Figs.~\ref{fig5}-\ref{fig10} for the specified parameter set. Please note that the trajectories are not mediated and they are all given for the same timespan of $150000$ timesteps in order to enable a comparison between different cases. Description of the results is given in terms of the standard deviation $\sqrt{\sigma _{\Phi i} ^2 h}$, further denoted by $\sigma _S$.

For the numerical procedure, $f_i^{(s)}$ is a zero mean Gaussian white noise of volatility $\sigma _f$ and variance $\sigma _f^2$, denoted $\mathcal{G}(0, \sigma _f )$. Its dimensionless counterpart, $\Phi_i^{(s)}$ is a zero mean Gaussian white noise of volatility $\sigma _\Phi$ and variance $\sigma _\Phi^2$, denoted $\mathcal{G}(0, \sigma _\Phi )$. Further, $dW_i$ is a zero mean Wiener process with volatility $\sigma _W$ and variance $\sigma _W^2$, a normal variable $\mathcal{N}(0, \sigma _W )$.

In the dimensional equation of motion with noise $f_i^{(s)}$, the variance $\langle f_i^{(s)} f_j^{(s)}\rangle = \sigma ^2_f$. When implementing the dimensionless equations, we take into account that $\langle \Phi_i^{(s)} \Phi_j^{(s)}\rangle = \sigma ^2_\Phi$ and the connection between the two variances is $\sigma _\Phi ^2 = \alpha ^{-4} r_0^4 \sigma _f^2 $, which for a proton in the St\"{o}rmer problem of the Earth is of the order $10^{-68}\sigma _f^2$.

In the Milstein procedure, the random number drawn at each timestep is from a distribution $\mathcal{N}(0,\sigma _W )$; note that $\sigma _W$ directly includes the value of the timestep $h=10^{-3}$.

A note regarding the appearance of escape trajectories: BM particle trajectories are just one realization of the Brownian path. When repeating the trajectory many times, i.e., given an ensemble of identically prepared particles allowed to follow a chaotic trajectory in a thermal bath, some of the particles in the trajectory will escape, although in a non-stochastic context they would be trapped. 

To calculate the (dimensionless) radiation appearing in the process of BM in the St\"{o}rmer problem, and such that the result may in principle be comparable to observations, one needs to mediate over many realizations of the stochastic process. As such, all radiation curves are mediated for $10^4$ statistically independent paths, that is to say that the random numbers $N_{1i}$ and $N_{2i}$ for trajectory $k$ are independent from those drawn for trajectory $k+l$; each of the paths has $150000$ timesteps: initial conditions are kept identical, but each path is subjected to statistically independent noise. The radiation curve is a mean of all these paths.

\subsubsection{Numerical results} 

In the case of the periodic motion, the combined effects of the presence of dissipative and stochastic forces is represented in Figs.~\ref{fig5}-\ref{fig7}. The effects of the increase of the dissipation coefficient, generally leading to a decrease in the radiation intensity, and periodic motion patterns are compensated by the presence of the stochastic force, leading to a slower decay of the motion. However, for large value of $\Gamma$ of the order of $\Gamma =10^{-2}$, even in the presence of the Brownian component in the motion, the radiation intensity tends to zero in the large time limit, but with a modified power distribution, as compared with the simple dissipative case. 

The effects of the random force can also be seen on the PSD of the process, which shows a significant difference with respect to the deterministic dissipative case, with the presence of two sharp peaks shifted, for small values of $\Gamma$, to higher values of $\omega$. However, for $\Gamma =10^{-2}$, the position of the main peak is again obtained for small values of $\omega$, with a second peak located at $\omega \approx 10^{-6}$. The presence of two peaks in the PSD is thus a common characteristic of both CDSP and SDSP.

The impact of the stochastic force on the chaotic motion in the CDSP is presented in Figs.~\ref{fig8}-\ref{fig10}. In this case there is a significant effect of the stochastic force on the motion, radiation, and statistical properties. For a small value of $\Gamma =10^{-4}$, the radiation pattern is characterized by the presence of several sharp peaks in the radiation intensity, with the first major peak occurring rather late during the dynamical evolution. 

The PSD function has a complex structure, indicated by the presence of multiple peaks, whose maximum values decrease with increasing $\omega$. With the increase of $\Gamma$, the radiation intensity and the PSD of the chaotic SDSP change drastically, and new and distinct patterns emerge. For $\Gamma =10^{-3}$ a single high intensity peak can be observed at large times as appearing in the emitted electromagnetic power, as shown in Fig.~\ref{fig9}. The PSD has also two peaks for small values of $\omega$, and it decreases for larger $\omega $ values. For $\Gamma =10^{-2}$, the maximum of the radiation intensity is reached at the beginning of the dynamical evolution, and the emitted power decreases for very small values in the large time limit. The evolution of the PSD function is characterized by the presence of two peaks, and a decrease towards zero for large values of $\omega$.                 

    \subsection{Escape rates}

In the deterministic problem, as long as the problem is specified, i.e., one knows the potential in which the motion occurs, the issue of escape is settled the moment the initial conditions are specified. This is not the case for the stochastic counterpart. It is possible that the individual random kicks a particle receives changes its trajectory type from bounded to escape. The question is then how often does this happen and what affects this rate?

Simulations were set up to calculate the escape rate: for a fixed set of parameters comprising of initial conditions, friction magnitude and noise magnitude, $N_{traj}=10^4$ different trajectories were analyzed to count how many are escape trajectories. The escape rate is then calculated as the percent of escape trajectories out of the total number of trajectories. The same approach was then taken for different sets of frictions and noise magnitudes. Based on physical considerations, it is expected that the escape rate decreases with increasing friction (for constant noise), and increases with increasing noise (for fixed friction). The purpose is to determine a functional form of the dependency of escape rate with these parameters.

\begin{figure*}
	\centering
	\includegraphics[width = 0.5\textwidth]{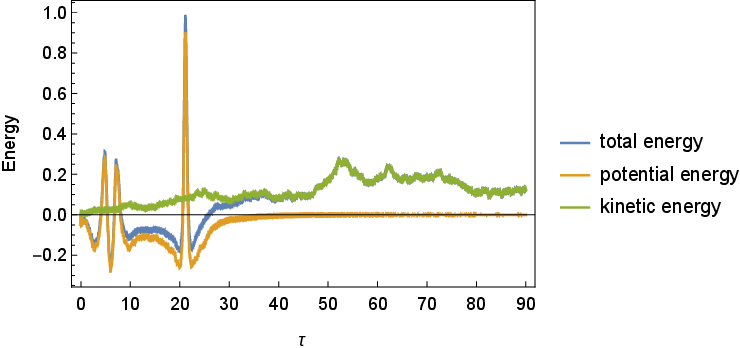}
	\includegraphics[width = 0.35\textwidth]{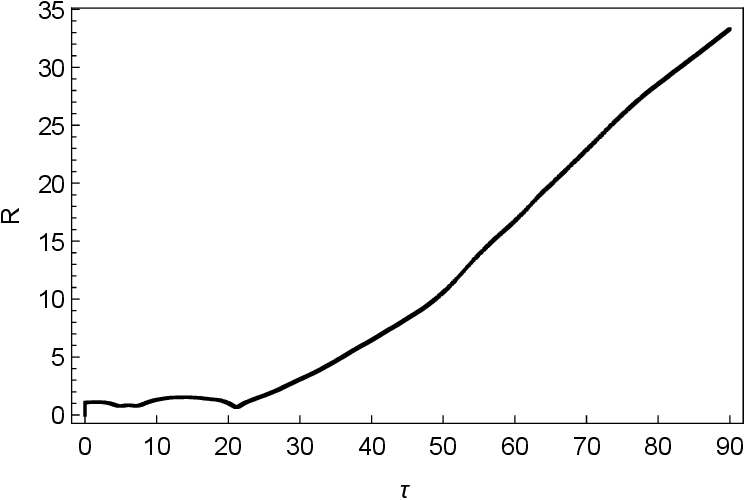}
	\caption{Brownian Motion in the Dissipative-Stochastic St\"{o}rmer Problem: energy (left) and distance (right) for a sample trajectory with $\vec{R}_0=\left(0.7,0.8,0\right)$ and $\vec{V}_0=\left(0.01,0.10,0.10\right)$, $h=0.001$, $L=150000$, $\sigma _S= 10^{-3}$, $\Gamma= 10^{-3} $.}\label{fig:escape-basis}
\end{figure*}

The decision of whether or not a trajectory is an escape trajectory had to be implemented in the code, as it is infeasible for a human operator to analyze the large number of trajectories generated. While the most elegant considerations for deciding escape are based on energy (Fig.~\ref{fig:escape-basis} left panel: the interplay between kinetic and potential), the decision tree that will give no false positives (i.e., a false decision that a trajectory is an escape one) is one based on the unbounded growth of distance $R = \sqrt{X^2+Y^2+Z^2}$ (Fig.~\ref{fig:escape-basis} right panel).

The dimensionless energy is $\overline{E}=E/\left(m\beta^3 r_0^2\right)$ and is given by the sum between kinetic and potential
\begin{equation}
	\overline{E} = \frac{1}{2}\left ( V_x^2+V_y^2+V_z^2\right) + \frac{1}{R^3}\left(V_xY-V_yX\right).
\end{equation}

\subsubsection{The numerical algorithm for the escape rate.} The algorithm $\mathcal{H}$ used is as follows: We ran each trajectory for $L=150000$ timesteps. We considered the last $L_1=10000$ steps of each trajectory and asked if this array is well fitted by a straight line distance vs. time. If the $R^2$ of the fit is $\geq 0.9$, then the trajectory is being counted an escape trajectory.

Hence, most completely stated, our results present the escape rate of an ensemble of $N_{traj}$ particles, followed in time for $L$ timesteps, where the escape verdict is given by the algorithm $\mathcal{H}$.

\begin{figure*}
	\centering
	\includegraphics[width = 0.45\textwidth]{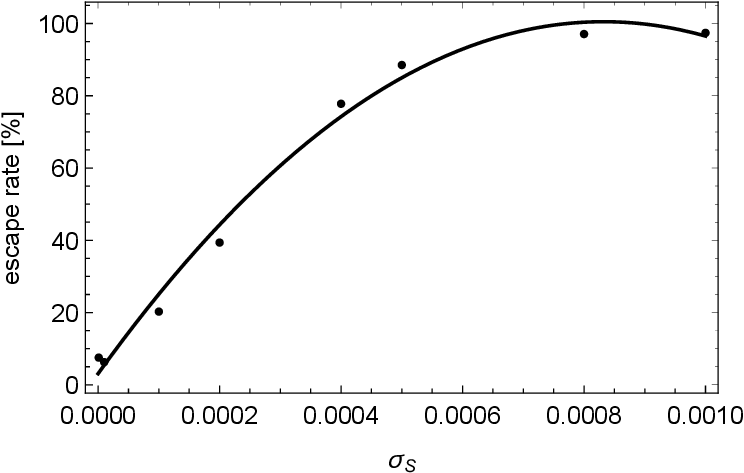}
		\includegraphics[width = 0.45\textwidth]{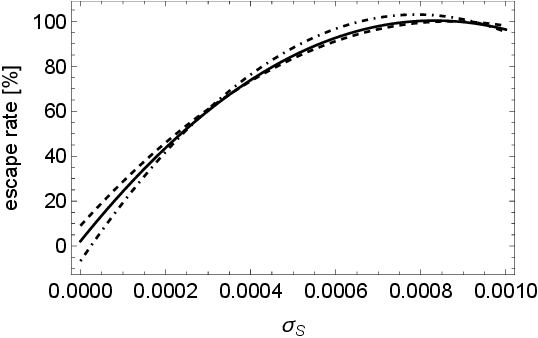}
	\caption{Stochastic-Dissipative St\"{o}rmer Problem escape rate as a function of noise magnitude for an ensemble of trajectories. Left panel: chaotic motion with $\vec{R}_0=\left(0.7,0.8,0\right)$ and $\vec{V}_0=\left(0.01,0.10,0.10\right)$.  Right panel: showing only the fitting functions for several cases. Full line: the chaotic case from the left, dot-dashed: the periodic motion and dashed: another chaotic motion with $\vec{R}_0=\left(0.7,0.8,0\right)$,  $\vec{V}_0=\left(0,0.10,0.10\right)$. For all curves, $h=0.001$, $L=150000$, $\Gamma= 10^{-3} $. Each ensemble has $N_{traj}$ realizations and $\sigma _S$ is the only parameter which varies between ensembles.}\label{fig:SDSP4-escapeRate-f1e-3}
\end{figure*}

The variation of the escape rate as a function of noise magnitude is given in Fig.~\ref{fig:SDSP4-escapeRate-f1e-3} for trajectories with initial conditions that would render CSP both periodic and chaotic. The escape rate depends on the noise and friction, and is not manifestly different even if otherwise the trajectory ensemble would have been made of periodic trajectories. 

The fitting function has the equation escape rate $= -1.40 \times 10^8 \sigma _S ^2 + 2.34 \times 10^5 \sigma _S + 3.09 $. 

\section{Conclusions}\label{sect3}

In the present paper we have considered an extension of the classical St\"{o}rmer problem, by generalizing the standard equations of motion of a charged particle in a dipolar magnetic field through the inclusion of a friction term, describing dissipation, and a stochastic term, which is the result of the presence of a randomly fluctuating force. Hence, the equation of motion of the particle takes the form of a Langevin type stochastic differential equation. 

The Langevin equation has been used for a long time for the description of the stochastic processes, and for their applications in astrophysics and astronomy, beginning with the classic study of Chandrasekhar \citep{Chand}.  The frictional term is usually assumed to be given by Stokes law, with the dissipative constant $\gamma $ given by $\gamma =6\pi a \eta$, where $a$ is the particle radius, and $\eta$ is the coefficient of viscosity of the medium surrounding the particle \citep{Chand}. 

From the point of view of the applications to the Earth magnetosphere, the classical St\"{o}rmer contains three important approximations.  First of all, the magnetic field of the Earth has a complex structure, including a strong quadrupolar component  \citep{Ti}, which is not
considered in the classical problem. The second important simplification is related to the assumption that
the dipole is not rotating. Moreover,  due to the rotation of the Earth a time periodic force must be added to the equations
of motion. Finally, as a third simplification, the St\"{o}rmer model ignores the energy losses through the radiation of the charged  particles (Bremsstrahlung), when moving under the influence of external forces. Moreover, the effects caused by the non-concentricity of the Earth and of its magnetic dipole are not taken into account. In our present approach, even if we neglect the complicated structure of the magnetic field near realistic astrophysical objects, as well as the rotation of the central object, the energy loss via various radiation mechanisms is taken into account through the inclusion of the dissipative force in the equation of motion. 

The magnetosphere of the Earth is strongly affected by the geomagnetic (or Solar) storms, which are severe disturbances caused by the interaction of the magnetosphere with  solar wind shock wave, or clouds of magnetic fields that interact with the magnetic field of the Earth \citep{Gon}. 

The most common causes of the geomagnetic storms are a Solar coronal mass ejection, or a co-rotating interaction region with  a high-speed stream of solar wind originating from a coronal hole. Hence, generally,  geomagnetic storms can be considered as the result of the random interaction between solar winds with the magnetic field of the Earth \citep{Zhao}.  

It turns out that the electric field of the solar wind, as well as its dynamic pressure are the main parameters that determine the intensity of an important geomagnetic storm. Sudden and strong  depletion in the equatorial ionospheric plasma density is also observed during magnetic storms, and it is called the Equatorial Plasma Bubble \citep{Yo}. From a physical point of view geomagnetic storms can be explained by the increase in the Earth's ring current \citep{Wa}.  The currents moving towards the west cause large‐scale fluctuations in the magnetic field of the Earth, which can be measured on the surface. 

 Other notable effects of the geomagnetic storms  are the changes in density of the ionosphere, like, for example, the increase in the mid‐to‐low latitudes of the densities of the total electron component \citep{Wa}. 
 
 All the above mentioned  effects are essentially random in their nature. Hence, we propose to model, at least in a first order of approximation, the effects of the geomagnetic storms via the mathematical formalism of the Stochastic-Dissipative St\"{o}rmer Problem.

While the properties of the trajectories of the particles of the St\"{o}rmer problems have been intensively studied, the characteristics of the electromagnetic radiation emitted by a particle in a magnetic dipole field seem to be less investigated.  The radiation from a normal star with a strong dipole magnetic field outside its surface, in which ultrarelativistic electrons are spiraling, was investigated in \cite{Th}. The polarization and intensity of the synchrotron component of the radiation was obtained. Other investigations of radiative properties of charged particles in dipole magnetic fields were performed in \cite{DR1,DR2,DR3}. In the present work we have also computed the radiation emitted by the charged particles in the dipole field, as well as the radiation spectrum.

The numerical algorithm presented in this paper starts from the analytical formulation of the Classical St\"{o}rmer Problem and it includes both noise and interactions with a heat bath. The resulting equations of motion can only be solved numerically. As a consequence, a number of numerical methods were aggregated in order to answer the physics questions stemming from the SP and most importantly from the desire to compare the SP with observations. The solution for two CSP cases was obtained and it was verified by the $K1-0$ method of \cite{gott2004} that one case was not chaotic and one case was chaotic. Trajectories, radiation patterns and PSD of the radiation patterns were produced for the CSP cases and also for the same cases when friction is present in the equation of motion. A multidimensional Milstein scheme \cite{Mi} was used to produce the solution of the equation of motion of the St\"{o}rmer particle in Brownian Motion. Radiation patterns were obtained by mediating over an ensemble of trajectories and the sensibility to physical parameters was studied. The PSDs for all these radiation patterns were produced.

We would like to point out that in all our numerical investigations we have adopted a value of the initial position of the particles given by $1.063\times r_0$, which in the case of the Earth, with $r_0=R_{\oplus}=6371$ km gives an initial position of around 400 km above the Earth surface. This is twice as high as the South Atlantic anomaly \cite{An}, and it is of the order of the average altitude of the International Space Station.

The numerical results reproduce the known physical behavior of the two types of trajectories in the CSP, the periodic and the chaotic one, both bounded. The novelty comes from considering these trajectories in the more realistic context of non-zero friction and interaction with a heat bath. As with all numerical endeavors, the volume of results is large and only a part of the results was explicitly included in the paper. 

Some general points to be made about {\it the novel results obtained throughout the paper} are

\begin{itemize}
	\item For {\it the periodic Classical St\"{o}rmer Problem (CSP)} cases, the PSD has a sharp peaked structure, as expected.  The trajectory and emitted radiation are identifiable periodic. It was found that if one adds friction (DCSP case), the PSD of the emitted energy looses its single peaked structure. A more complex structure appears in the PSD for a larger period. The CSP period of the system does not change significantly if friction is included, i.e., the corresponding peak does not move in the PSD, but its amplitude decreases by one order of magnitude.
	\item For {\it the chaotic Classical St\"{o}rmer Problem  (CSP)} cases, the PSD has more peaks, with a stronger peak at the period of the (periodic) CSP motion. This period is a characteristic of the system of equations. The trajectories and the PSDs of two different initial conditions for the chaotic trajectories look different. The emitted radiation has a somewhat similar pattern, but the amplitudes differ by one order of magnitude.
	\item When {\it friction} is added to the CSP chaotic cases ({\it the Dissipative Classical St\"{o}rmer Problem (DCSP)}),  it is noted that as the friction increases, the peak structure of the PSD decreases in complexity. The main period as extracted from the PSD increases as the friction increases, i.e., the peaks moves to the left on the positive frequency axes.
	\item For {\it the periodic case, with both friction and noise (the Stochastic-Dissipative St\"{o}rmer Problem (SDSP))}, the trajectory keeps its periodic look although the motion is clearly noisy. The PSD retains its peaked aspect, and shows a peak at the period of the corresponding CSP. The radiation pattern does not show significant noise presence.
	\item For the {\it chaotic cases, with both friction and noise (SDSP)}, the trajectories show a marked noisy component, although a balance between periodicity-friction and noise may be attained in the parameter space. The radiation pattern changes significantly with respect to SDSP periodic. The PSD also shows a more complex structure as a result of including noise in the equations of motion. 
	\item The {\it escape rate} of a St\"{o}rmer Brownian particle in the SDSP was calculated, and it was shown that it increases with increasing noise amplitude as a second degree polynomial.
\end{itemize}

An interesting question is if the algorithm considered in the present work could be used  in the differential diagnosis of data.  We propose that it can, in a setting including observational data analysis and knowledge about at least some of the parameters describing the system of charged particles emitting the observed radiation.

Despite the physical simplifications it contains, the St\"{o}rmer problem, and its different versions, represent an efficient and successful way in analyzing the dynamics of charged particles in a dipole magnetic field. Even its first formulation, the classical St\"{o}rmer problem, is of great mathematical complexity, with the motion of particles ranging from simple oscillatory behaviors to chaotic regimes. Its generalization to the case of the Stochastic-Dissipative St\"{o}rmer problem may open a new perspective on the complex dynamics of charged particles in an astrophysical environment.  

\section*{Acknowledgements}

We would like to thank the anonymous reviewers for comments and suggestions that helped us to improve our manuscript. 

\section*{Conflict of interest}

The authors declare no conﬂict of interest.

\section*{Keywords}

St\"{o}rmer problem, dissipation, random forces,  Lorentz-Langevin equation, radiation emission





\end{document}